# Characterizing Online Toxicity During the 2022 Mpox Outbreak: A Computational Analysis of Topical and Network Dynamics

Lizhou Fan, Lingyao Li, Libby Hemphill
*University of Michigan, School of Information*
lizhouf@umich.edu, lingyaol@umich.edu, libbyh@umich.edu

## Abstract

**Background:** Online toxicity, encompassing behaviors such as harassment, bullying, hate speech, and the dissemination of misinformation, has become a pressing social concern in the digital age. Its prevalence intensifies during periods of social crises and unrest, eroding the sense of safety and community. Such toxic environments can adversely impact the mental well-being of those exposed and further deepen societal divisions and polarization. The 2022 Mpox outbreak, initially termed "Monkeypox" but subsequently renamed to mitigate associated stigmas and societal concerns, serves as a poignant backdrop to this issue.
**Objective:** In this research, we undertake a comprehensive analysis of the toxic online discourse surrounding the 2022 Mpox outbreak. Our objective is to dissect its origins, characterize its nature and content, trace its dissemination patterns, and assess its broader societal implications, with the goal of providing insights that can inform strategies to mitigate such toxicity in future crises.
**Methods:** We collected more than 1.6 million unique tweets and analyzed them from five dimensions, including context, extent, content, speaker, and intent. Utilizing BERT-based topic modeling and social network community clustering, we delineated the toxic dynamics on Twitter.
**Results:** By categorizing topics, we identified five high-level categories in the toxic online discourse on Twitter, including disease (46.6%), health policy and healthcare (19.3%), homophobia (23.9%), politics (6.0%), and racism (4.1%). Across these categories, users displayed negativity or controversial views on the Mpox outbreak, highlighting the escalating political tensions and the weaponization of stigma during this infodemic. Through the toxicity diffusion networks of mentions (17,437 vertices with 3,628 clusters), retweets (59,749 vertices with 3,015 clusters), and the top users with the highest in-degree centrality, we found that retweets of toxic content were widespread, while influential users rarely engaged with or countered this toxicity through retweets.
**Conclusions:** Our study introduces a comprehensive workflow that combines topical and network analyses to decode emerging social issues during crises. By tracking topical dynamics, we can track the changing popularity of toxic content online, providing a better understanding of societal challenges. Network dynamics spotlight key social media influencers and their intents, indicating that addressing these central figures in toxic discourse can enhance crisis communication and inform policy-making.

# Introduction

The 2022 Mpox outbreak was first reported in several countries in Europe and quickly became a global health crisis [1]. Mpox is a viral illness that is transmittable from animals to humans or between humans through contact with blood or bodily fluids [2]. This disease was originally named "Monkeypox" and was later renamed to reduce stigma and other issues during the 2022 outbreak [3]. This outbreak was characterized by a large number of cases and a high rate of transmission, which posed a significant threat to public health globally. Health authorities and public health organizations were quick to respond to the outbreak, implementing measures to control the spread of the disease, providing health care and support to those affected, and making vaccination efforts to protect those who were not yet infected [4–6].

During the Mpox outbreak, social media platforms were used for public health communication and information sharing about the disease, its spread, and people's feelings about it. This led to both positive and negative consequences, with accurate information being shared alongside misinformation and toxic comments [7]. Online toxicity is prevalent during health crises, with many individuals spreading misinformation, fear, and hate [8,9]. This can undermine public health communication efforts and create confusion and fear among the public. Additionally, online toxicity disproportionately impacts historically marginalized communities, exacerbating existing health disparities and making it difficult for these communities to access accurate and trustworthy information during a crisis. Homophobia and racism were common in Mpox discussions [7].

Understanding the online toxic discourse during the 2022 Mpox outbreak is crucial for several reasons. First and foremost, studying toxicity in Mpox discussions helps identify the factors contributing to the spread of misinformation, fear, and panic, which can exacerbate the public's response to the outbreak. Second, by analyzing the motives and patterns behind such toxic behavior, public health officials and researchers can develop effective communication strategies to counteract negativity and promote accurate information. Lastly, understanding the prevalence and impact of toxic discourse allows us to explore the broader implications of online behavior on societal discourse and public opinion formation during health crises. To achieve this understanding, we chronicled original *tweets*, the online posts on the Twitter social media platform, from May to October 2022. We then use the Perspective API [10,11] to identify the *toxic tweets*, the rude, disrespectful, or unreasonable comments on Twitter that are likely to encourage individuals to leave conversations. We analyzed the *toxic tweets* to answer the following research questions:

- RQ1, ***aboutness*****:** What are the *toxic tweets* about and how do they change through time?
- RQ2, ***diffusion*****:** How do *toxic tweets* spread through online social networks?

The *aboutness* of *toxic tweets* highlights the connection between the evolution of toxic discourse topics and the motivations of the people posting these tweets. This analysis helps us understand the thoughts and concerns of the ongoing public health emergency for a significant portion of the public, including hundreds of millions of people who are active on Twitter's social network.[1] The *diffusion* of *toxic tweets* is summarized as the retweets and mentions networks.

---

[1] As of September 6, 2023, there are about 330 million monthly active Twitter users [12].

The retweets network can inform who initiated or distributed toxic comments, while the mentions network shows who was frequently mentioned by other users and thus should be aware of toxicity dissemination. In particular, we followed the analytical framework of five dimensions, namely *Context*, *Extent*, *Content*, *Speaker*, and *Intent*. The five dimensions in characterizing toxic discourse, as adapted from the Rabat Plan of Action [13], are crucial for understanding the relationship between online toxicity and public health policy.

Our study aims to unravel the complex dynamics of online toxicity during the 2022 Mpox outbreak by employing advanced computational techniques. Specifically, we aim to identify the thematic structures, the *aboutness*, and the network behaviors, the *diffusion*, that perpetuate toxic discourse, to understand how such narratives spread and the role of influential network actors in this process. Ultimately, our study seeks to offer actionable insights that can help design effective interventions to mitigate online toxicity in future public health emergencies and other crisis situations.

## Online toxicity on social media

Online toxicity on social media refers to rude, aggressive, and degrading attitudes and behaviors, which are exhibited in various forms, including harassment, bullying, or even the spread of hate speech and misinformation [14,15]. One of the primary causes of toxicity on social media is the anonymity and physical disconnect provided by the cyberspaces of online platforms. The online disinhibition effect magnifies the toxicity and facilitates the implementation of toxic ideas in daily life [16], which makes online toxicity a useful tool to anticipate extremes in public opinions and social dynamics. At the same time, online toxicity is easily contagious and can propagate quickly through social networks [17], where algorithms deployed by online social media platforms can contribute to the spread of toxic content by amplifying it and showing it to a larger audience [18].

Consequently, toxicity on social media can negatively impact the mental health and well-being of individuals, contributing to anxiety, depression, and low self-esteem, among other issues [19,20]. The aggravating toxicity on social media can exacerbate the culture of hate and division, causing harm to marginalized communities and making it difficult for minority individuals to engage in meaningful and productive online discourse and face-to-face activities [21].

There are recent efforts in academia and industry that seek to facilitate understanding of online toxicity and implement detection and moderation through socio-technical approaches. Researchers have gained insights of online toxicity from different perspectives. [22] quantify toxicity and verbal violence through crowdsourcing, which can be useful for moderation of toxic contents. [23,24] both detect triggers of online toxicity and better understand the causes of toxic discussions from topical and sentiment shifts in interactions. [20] find meaningful context of online conversations can help highlight or exonerate purported toxicity.

Benefiting from increasingly less expensive cloud storage and computing, social media platforms have also started developing and deploying online toxicity moderation applications. Perspective API developed by Jigsaw and Google's Counter Abuse Technology team provides free access to toxic content detectors that aims at enabling healthy conversations and reducing toxicity and abusive behavior [10,11]. The OpenAI moderation endpoint, similarly, is a tool for checking content's compliance with OpenAI's content policy, including the prohibition of the generation of hateful, harassing, or violent content [25].

In addition to analysis and moderation of toxic contents, other research calls for public engagement to tackle online toxicity, including encouraging individual responsibility and positive behaviors, as well as raising awareness and education [26,27]. Despite the efforts to study online toxicity, it remains hard to massively eradicate the impact of existing toxicity and the spread of new toxicity. It is still imminent and meaningful to keep track, enhance comprehension, and intensify awareness of online toxicity on social media.

## Health crisis communications on social media

Health crisis communications on social media refers to the use of social media platforms to disseminate information and communicate during public health emergencies and crises [28]. Social media has become an important tool for public health organizations and governments to communicate with the public during times of crisis and emergency, as it can reach a large and diverse audience quickly and effectively, engaging with the public and gathering feedback [29,30]. Individual users of social media can also widely read others' opinions about a health crisis, freely express their feelings, and receive timely feedback [31].

However, there are also several challenges associated with using social media for health crisis communications. One of the main challenges is health-related online toxicity on social media that can lead to confusion and fear among the public. For instance, during the COVID-19 pandemic, there was a surge in misinformation, conspiracy theories, and discriminatory remarks which not only hindered effective public health responses but also fueled stigma, discrimination, and even violence against certain groups [32,33]. Additionally, the volume of information and number of sources on social media can be overwhelming for the public, making it difficult for them to discern what information is reliable and relevant [34]. Without proper planning and intervention, information on social media can negatively influence health crisis communications and even result in infodemics [35,36]. Thus, public health organizations and governments should develop clear and consistent communication strategies for monitoring and responding to social media and provide accurate and trustworthy information to the public.

There are several guidelines that are helpful for public health organizations and governments to refer to. The Center for Risk Communication suggest six best practices in public health risk and crisis communication, including (1) accept and involve stakeholders as legitimate partners, (2) listen to people, (3) be truthful, honest, frank, and open, (4) coordinate, collaborate, and partner with other credible sources, (5) meet the needs of the media, and (6) communicate clearly and with compassion [37]. Recent research on COVID-19 also suggests providing relevant, accurate, and sensitive information to key public groups to minimize communication noise and guide desirable coordinated actions [34]. While these principles are carefully written, it remains challenging to implement them in practice, especially due to the complexity of social media's role in health crisis communications.

# Methods

We retrieved a large corpus of toxic online discourse on Twitter and applied computational methods for analysis. In this section, we describe data and methods used in the paper (**Figure 1**). We first introduce the data retrieval process with *extent* as a content relevance filter, followed by preliminary analysis of *context*. We then provide the details of methods for characterizing topical and network dynamics, supporting the comprehensive analysis of the *content*, *speaker*, and *intent* (of social media users).

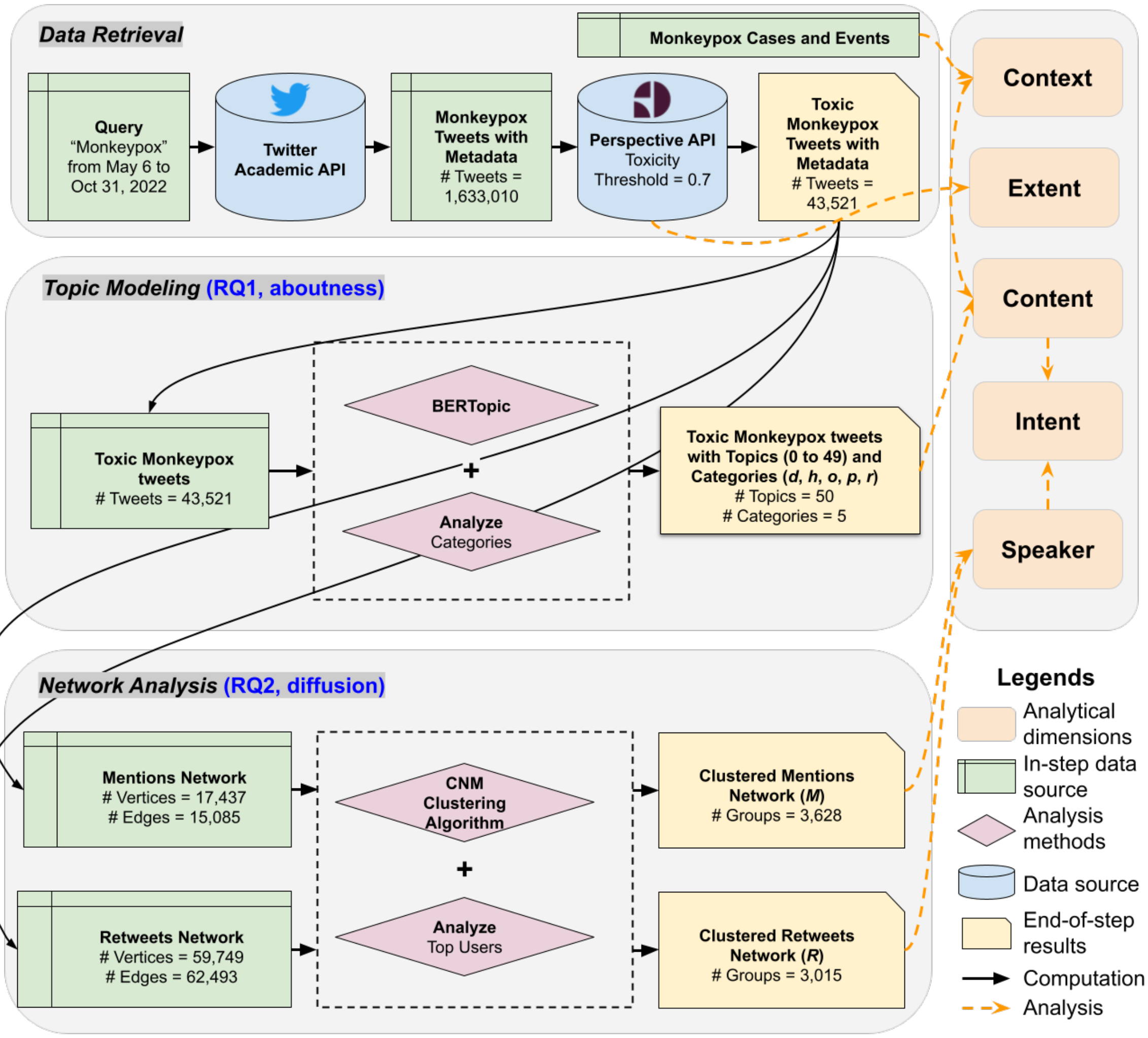

**Figure 1.** Workflow overview. We first used external APIs to retrieve and identify toxic Twitter data and Mpox context data. We then applied topic modeling and network analysis methods to categorize temporal topical dynamics and to cluster network communities. We also mapped the results in each step to the analytical framework.

# Data

In this section, we demonstrate the data collection process and the context of volume peaks.

## Retrieval of toxic tweets

To chronicle online discourse on the 2022 Mpox outbreak, we used the Twitter Academic API

[38][2] to query tweets with the keyword "monkeypox"[3]. The collection of 1,633,010 unique English-language tweets started on May 6, 2022, when the initial cluster of cases was found in the UK [39], and ended on October 31, 2022.

We then applied the Perspective API developed by Jigsaw and Google's Counter Abuse Technology team to identify *toxic tweets* [11]. They define *toxicity* as a rude, disrespectful, or unreasonable comment that is likely to disengage others' participation [11], especially those who are targeted by such toxicity. We adopted their definition of toxicity.

The Perspective API assigns each text submitted a probability score that corresponds to the proportion of people who would consider the text toxic. While choosing an appropriate threshold depends on the specific use case, the Perspective API team suggests researchers experiment with a threshold between 0.7 and 0.9 to classify toxicity [11]. Based on our dataset, we observed that a tweet with a score higher or equal to 0.7 generally implied toxicity and therefore chose 0.7 as the threshold to identify toxicity. When we limited our dataset to tweets that receive 0.7 or higher scores from Perspective API, 43,521 *toxic tweets* remained.[4]

## Temporal context and hashtags

As **Figure 2** shows, we observed two significant peaks in the volume of toxic public discourse on Twitter related to the Mpox outbreak. The first peak occurred from mid to late May, with a daily high of over 1,200 toxic tweets. This spike in toxic discourse coincided with the reporting of the initial cases of the 2022 Mpox outbreak [40]. Notably, this peak preceded the World Health Organization's (WHO) public acknowledgement of their tracking efforts of the global disease development of Mpox by approximately one month.

The second peak in toxic discourse was observed from late July to early August. This surge, which saw a daily high of around 2,200 toxic tweets, nearly doubled the volume of the first peak. This increase followed two significant events: the WHO declared the Mpox outbreak a global health emergency on July 23, and the CDC designated Mpox a nationally notifiable condition on July 27 [1,2].

It is important to note that numerous Mpox-related reports, particularly those with social and political implications, were released around and following the peak of new cases and the two peaks in toxic discourse volume. For example, on August 9, local police in the Washington, D.C. investigated an assault on two gay men as a hate crime; anti-gay rhetoric and references to Mpox were used during the assault [41]. Then on September 14 and October 5, there were reports on regional and racial disparities in health care (Mpox vaccine equity) and health outcomes (Mpox rates) [42,43]. These incidents underscore the profound real-world consequences of misinformation and toxic discourse, emphasizing the need for accurate and responsible communication on Mpox and related issues.

---

[2] Note that Twitter Academic API service was discontinued as of mid-2023, while the access to Twitter data is still available with a cost [38].

[3] Some relevant but less frequently used words, for example, "Monkey pox" or "Mpox", are not included for query simplicity and API efficiency. As such, the name change does not influence our data collection.

[4] The Twitter ID, toxicity scores, and relevant metadata of these toxic tweets are available upon reasonable request. There are 43,521 unique non-retweet English tweets archived, while 27 duplicate tweets are removed for network analysis in later sections.

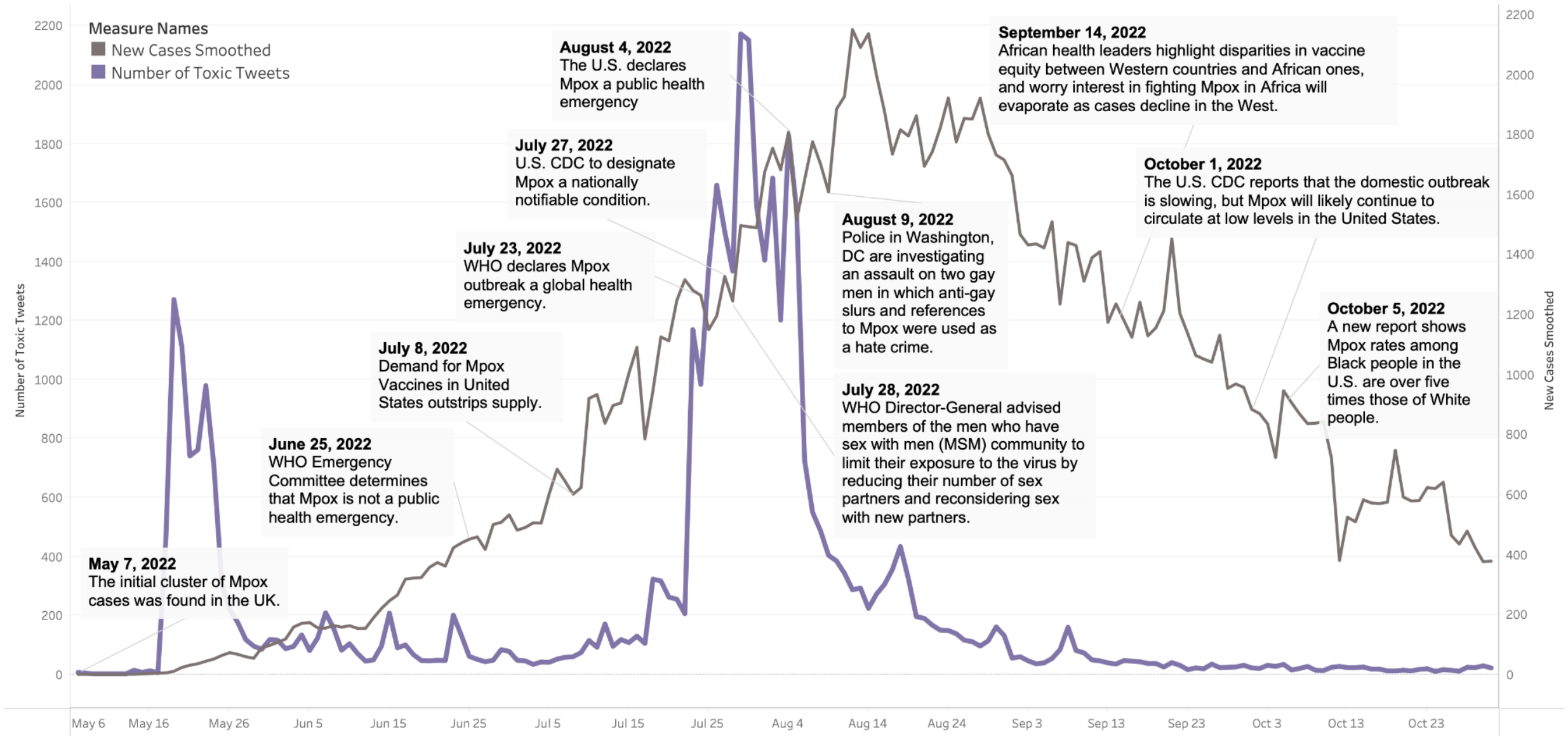

**Figure 2**. Development of the 2022 Mpox Outbreak [1] and the volume of toxic tweets

**Table 1** shows the summary of the top 20 hashtags in the toxic tweets, which highlights the frequent topics and provides a simplified overview of toxic online discourse on Mpox. There is a diverse use of hashtags on different aspects of the Mpox outbreak, including health emergency (e.g., “pandemic” and “covid19”), healthcare (e.g., “vaccine”), and population (e.g., “lgbtq” and “gay”). These hashtags demonstrate the variety of topics in toxic tweets, which occur in different aspects of the online discourse.

We also observe different types of entities in the hashtags, relating to health administration organizations (e.g., “who” and “cdc”), impactful individuals (e.g., “biden” and “trump”), and news agencies (e.g., “foxnews”). These hashtags indicate that further identifying users in this Twitter collection can help understand different stakeholders and participants in health and crisis information diffusion. Meanwhile, except for the query word “monkeypox”, all other hashtags are used less than 107 times. The variety of hashtags used, and their low individual frequencies, suggest that hashtags should not be the only source of identifying topics for this corpus. Alternative methods, such as topic modeling directly from the tweets, can be useful.

**Table 1.** Top 20 hashtags used in toxic tweets

| Hashtag | Count | Hashtag | Count |
|---|---|---|---|
| monkeypox | 2889 | cdc | 30 |
| monkeypoxvirus | 106 | billgatesbioterrorist | 27 |
| covid19 | 102 | lgbt | 23 |
| covid | 96 | trump | 22 |
| texasschoolmassacre | 77 | pandemic | 22 |
| gay | 76 | vaccine | 21 |
| who | 60 | monkeypoxalypse | 21 |
| lgbtq | 50 | foxnews | 20 |
| aids | 33 | pride | 19 |
| biden | 30 | idiots | 18 |

# Analytical Methods

In this section, we provide the details of the analytical framework for online toxicity in our analysis. We also illustrate the characterization of topical and network patterns in the toxic tweets on Mpox. We first demonstrate the technical details of topic modeling. We then discuss social network analysis methods for analyzing toxic information diffusion.

## Adaptation of an analytical framework of hate speech analysis for online toxicity analysis

We leveraged the hate speech analysis by the Rabat Plan of Action [13,44] and adapted it to the online toxicity context. Our adapted analytical framework includes the five dimensions as follows:

- *Context*: the social and public health landscape behind toxic online discourse, including what *events* co-occur during the discourse and how *public health metrics* change relative to the volume and content trends;
- *Extent*: the severity to which the message can be considered *abusive or harmful* to the targeted group, which can be assessed with a score from 0 to 1;
- *Content*: the semantic summary of toxic online discourse, which reveals *attributes of the targeted group* (e.g., vulnerability, political representation, and social construct) and the discourse's *co-occurrence with other narratives* that are dominant in toxic discourse (i.e., major semantic clusters in the corpus);
- *Speaker*: the status of the social media user who posts toxic content, which can influence the *dissemination quantity* (indicated by network metrics) and *quality* (depending on the user's credibility, influence, and capacity);
- *Intent*: the assumed high level summary of *objectives* and *intended audience* for creating and spreading toxic content.

These five dimensions are fundamental for our analysis and can benefit different stakeholders. Analyzing the context of toxic discourse can help policymakers identify key events or trends that may be contributing to the proliferation of harmful messages, enabling them to address misinformation and foster a more supportive public health environment. Assessing the extent of toxic messages can help public health officials allocate resources and target interventions to counter the most severe cases of online abuse. Evaluating the content of toxic discourse reveals the attributes of targeted groups and dominant narratives, which can inform the development of tailored public health campaigns and interventions. Examining the speaker dimension provides insights into the dissemination of toxic content, allowing officials to monitor influential sources and mitigate their impact. Finally, understanding the intent behind toxic content can help public health policymakers craft strategies to engage with diverse audiences and counteract the harmful consequences of such discourse. By examining these five dimensions collectively, public health officials can gain a comprehensive understanding of the online toxic landscape during health crises, allowing them to devise timely and effective policy interventions. **Table 2** further compares these five dimensions for analyzing toxic information diffusion with the original hate speech analysis framework by the Rabat Plan of Action [13].

**Table 2. A comparison of analytical frameworks**

| Dimension | Original definition | Definition for online toxicity in health crises | Analysis of adaptation |
|---|---|---|---|
| **Context** | The social, cultural, and political landscape where the target of the hate speech is vulnerable. | The social and public health landscape behind online toxicity, including what *events* co-occur during the discourse and how *public health metrics* change relative to the volume and content trends. | We limit the scope to social and public health and refocus the context from only the targeted group to the health crisis-related sociality. |
| **Extent** | The magnitude of the dissemination efforts, or the extent of the hate speech act. | The severity to which the message can be considered abusive or harmful to the targeted group, which can be assessed with a score from 0 to 1. | We measure the semantic intensity instead of the diffusion extent. We used Perspective API [11] to score toxicity. Diffusion is analyzed with Speaker in our definition. |
| **Content** | Content and form, including, provocative degree or aggressiveness of the message, form taken by the expression, directness, call to action degree, correlation with other dominant hate narratives, and legal status. | The semantic summary of toxic online discourse, which reveals attributes of the targeted group (e.g., vulnerability, political representation, and social construct) and the discourse' co-occurrence with other narratives that are dominant in toxic discourse (i.e., major semantic clusters in the corpus) | We keep two relevant subdimensions from the original definition. |
| **Speaker** | The influence the | The status of the social | We summarize the |
| | speaker has on the audience to which the message has been presented, including status, capacity, credibility, and influence on the targeted group. | media user who posts toxic content, which can influence the dissemination quantity (indicated by network metrics) and quality (depending on the user's credibility, influence, and capacity) | original definition into two aspects, namely quantity and quality, which are about the speaker's influence. |
| **Intent** | Intent of the speaker is estimated from past actions, reactions after promoting the hate message, probable objectives, and the intended audience. | The assumed high level summary of objectives and intended audience for creating and spreading toxic content | We direct use the original definition in our context. |
| **Likelihood of immediate actions** | The likelihood of the speech act generating a situation that represents a clear and immediate danger to the targeted social group, which is useful to evaluate as being sufficiently extreme to require a criminal investigation of censorship from state institutions. | Not applicable. | Our purpose is not to intervene in online toxicity as independent researchers, but rather to identify and analyze them. Thus, the above five dimensions are sufficient. |

## Modeling topics in toxic online discourse

Topic modeling is a method for detecting and analyzing latent semantic topics from large volumes of unstructured text data. It assumes that each text document (e.g., a tweet) is a combination of multiple latent topics, where each topic is represented by a probability distribution of words, while representing topics by grouping together words that have similar meanings based on their probability distributions [45,46]. Topic modeling identifies groups of words or vectors that appear together, and those groups are referred to as "topics". They are not necessarily topics in the colloquial sense of a "subject" or "theme". Identifying the content themes within and across topics requires manual inspection of the topics produced by the model. We refer to this step as "categorizing" and manually identified 5 themes ("categories") that capture all 50 topics. Topic modeling and other semantic presentation methods have been used as a big data analysis tool in a variety of fields of research, including social media studies, health informatics, and crisis informatics [47–49].

With the development of deep learning techniques, for example, transformers [50] and BERT [51], recent topic modeling methods take advantage of the embedding-based approach that better represents semantic relationships among words. These algorithms approach topic modeling as a clustering task and provide flexible language representation and text mining options [52,53]. As **Figure 3** shows, our study follows this state-of-the-art development in topic modeling. We implemented a human-computer hybrid methodology sequence including both computational steps (in purple frames) and a human step (in a gray frame) for modeling topics in toxic online discourse during the Mpox outbreak. We followed the default steps and setting in BERTopic [53], a neural topic modeling method with a class-based TF-IDF procedure. We also extended the standard procedure by adding the preprocessing and the categorizing steps.

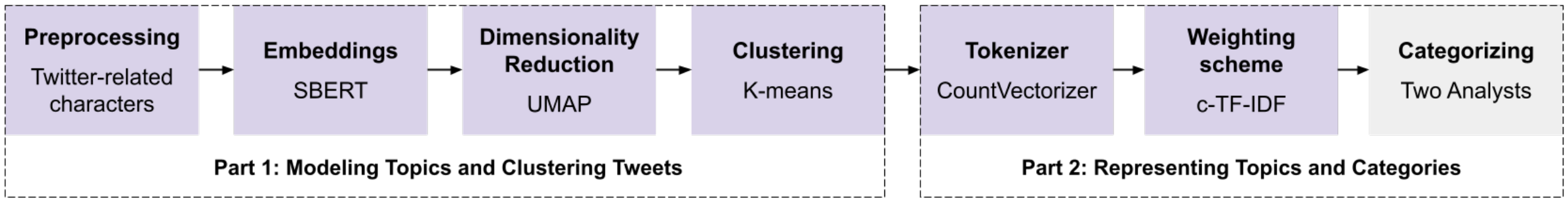

**Figure 3**. An extended sequence of steps with BERTopic

The first part of the BERTopic methodology sequence is modeling topics and clustering tweets. We started with preprocessing the toxic Mpox tweets by removing Twitter-related characters, including “@” and “RT” marks for social media networks and links starting with “http” for external web information. We then used Sentence-BERT (SBERT), a transformer-based pre-trained NLP model, to derive semantically meaningful sentence embeddings for each of the cleaned tweets [54]. In particular, we used the SBERT Python package and based on the pre-trained model ‘all-MiniLM-L6-v2’ [55], which enables clustering and semantic searching by mapping search tweets to a 384-dimensional vector space and works well for semantic similarity tasks. To better handle the high dimensional tweet vectors for clustering, we implemented a dimensionality reduction technique (UMAP) [56]. UMAP helps cluster models handle dimensionality [57] while maintaining a dataset's local and global structure. This feature of UMAP is important for constructing topic models, which depend on word vectors’ structural similarities.

We then used the scikit-learn implementation of Lloyd’s K-Means clustering algorithm to group similar sentences’ embedding vectors into topics [58]. We used K-Means because it ensures every vector is clustered into a topic and allows choosing the number clusters with experiments. It first initiates centroids in the 384 dimension vector space and randomly assigns a centroid to each of the 43,521 vectors. It then uses Euclidean distance [59] to update the centroid assignments recursively, which stops when the centroid assignments no longer update. We experimented with three different numbers of clusters: 30, 50, and 100. 30 clusters produced clusters with a mix of topics and prevented us from disambiguating them. 100 clusters produced sparse clusters in which we needed to manually combine many topics. We settled on 50 clusters as a compromise[5].

The second part of the sequence is representing topics and categories. We first tokenized topics using the count vectorizer in the scikit-learn Python package, which performs cluster-level (topic-level) bag-of-words representation that calculate and vectorize the frequency of each word in each cluster [61]. Here, we obtained word frequency vectors for each topic for representation purposes (i.e., extracting keywords in each topic). This is fundamentally different from bag-of-words topic modeling, which uses corpus-level frequencies for creating topics. We then used class-based term frequency-inverse document frequency (c-TF-IDF) to extract the difference of topical keywords, which help distinguish among the clusters. After converting each cluster (topic) into a single document, we extracted the frequency of word $x$ in class $c$ [53]. In c-TF-IDF, we then had the importance score per word in each class:

$$W_{X,C} = |tf_{X,C}| \times log\left(1 + \frac{A}{f_X}\right) \tag{1}$$

where $tf_{X,C}$ is the frequency of word $x$ in class $c$, $f_X$ is the frequency of word $x$ across all classes, and $A$ is the average number of words per class. In this way, we are able to represent topics with the unique and frequent words as the keywords.

[5] While the sentence embeddings remain the same, the clustering results can slightly vary with different random seed initiated in the background. As there are only trivial differences among the clusters of sentences, we pick one of the results for our analysis. We share the model on hugging face [60]. Due to Twitter’s content sharing restrictions, the trained embeddings are only available upon reasonable request.

We then characterized the 50 topics based on the keywords and original tweets. After reviewing related literature, iterative refining categories, and labeling samples to study each category, we annotated each topic with the following five categories: Disease (*D*), Health Policy and Healthcare (*F*), Homophobia (*O*), Politics (*P*), and Racism (*R*).

## Measuring user influence in the toxic tweet networks

Understanding how toxic information spreads on social media during public health crises is critical. Relationships, or user interactions, in social networks are often used to facilitate understanding of information diffusion in infodemiology [62]. We focused on two types of relationships on Twitter: mentions and retweets. A mention (i.e., @username) is a tweet that quotes another user's name in the text. The user who is mentioned will receive a notification from Twitter. A retweet is a reposting of a tweet that starts with "RT @username" [63]. We calculated three measures of influence in the network, in-degree centrality, out-degree centrality, and betweenness centrality, with close degree-centrality as a representative metric. Degree centrality refers to the number of edges a vertex has to other vertices and it defines three types of centrality [64]. In our study, we particularly focus on (1) in-degree centrality, which measures the number of incoming connections a node has, indicating its popularity or influence within the network, and (2) betweenness centrality, which assesses the extent to which a node acts as a bridge along the shortest path between other nodes. Given a network $G = (V, E)$ with $V$ vertices and $E$ edges (defined as $deg(v)$), in-degree can be computed as [64],

$$deg_I(v) = N_v \tag{2}$$

where $N_v$ denotes the total number of all the incoming edges into vertex $v$. The betweenness of vertex $v$ in a network is the fraction of all shortest paths between every pair of other vertices $(s, t)$ that pass through vertex $v$. This is computed in three steps: (1) for each pair of vertices $(s, t)$, compute the shortest path between them, (2) for each pair of vertices $(s, t)$, determine the fraction of the shortest paths that pass through the vertex $v$, and (3) sum the fraction over all pairs of vertices $(s, t)$. Then, the betweenness of vertex is calculated as [64],

$$deg_B(v) = \sum_{s \neq t \neq v \in V} \frac{\sigma_{st}(v)}{\sigma_{st}} \tag{3}$$

where $\sigma_{st}$ represents the total number of shortest paths from vertex $s$ to vertex $t$, $\sigma_{st}(v)$ is the number of those shortest paths that pass through vertex $v$. For more details of network analysis and metric calculation, please refer to Appendix II and **Figure 8**.

Next, we applied a tool called NodeXL to generate the social network [65]. NodeXL is a visualization tool for social network analysis that is implemented as an add-on in Microsoft Excel [65]. We further applied the Clauset, Newman, and Moore (CNM) algorithm [66] to investigate the social network in communicating toxicity on Twitter. The CNM algorithm infers the community structure from network topology that works by optimizing the modularity. It also provides insights into how vertices in social networks function and affect each other. One issue addressed by the CNM algorithm is to understand opinion leaders (e.g., opinion leaders have high in-degree centrality) and distributors (e.g., distributors have high betweenness centrality) in disseminating information.

As such, we generated social networks using mentions and retweets, respectively, and investigated the following metrics in each network. The statistics for social networks *M* and *R* are summarized in **Table 3** with the following attributes:

- *Vertices* – Twitter users in the social network;
- *Edges* – relationships between two Twitter users (i.e., retweet and mention);
- *Duplicated edges* – mention or retweet multiple times between two same Twitter users;
- *Self-loops* – users mention or retweet their own tweets that form self-loops;
- *Connected components* – a set of users in the network that are linked to each other by edges (i.e., clusters in the social network;
- *Geodesic distance* – the length of the number of edges of the shortest path between two Twitter users (i.e., two vertices in the network).

**Table 3**. Social network statistics for mentions and retweets networks.

| Network metric | Mentions network (*M*) | Retweets network (*R*) |
|---|---|---|
| Network type | Directed | Directed |
| Vertices | 17,437 | 59,749 |
| Total edges | 15,085 | 62,493 |
| Duplicated edges | 750 | 1,663 |
| Unique edges | 14,335 | 60,830 |
| Connected components | 3,628 | 3,015 |
| Self-loops | 5 | 375 |
| Max. geodesic distance | 26 | 21 |
| Avg. geodesic distance | 8.041 | 5.336 |

# Results

## Topic modeling and categorization

In this section, we report the topic modeling and categorization results, including the overall composition, temporal patterns, and representative tweets in each category. We summarize the 50 topics results into five *toxicity categories*, i.e. the toxicity about five topical discourse, including Disease (46.6%, Category *D*), Health Policy and Healthcare (19.3%, Category *H*), Homophobia (23.9%, Category *O*), Politics (6.0%, Category *P*), and Racism (4.1%, Category *R*).

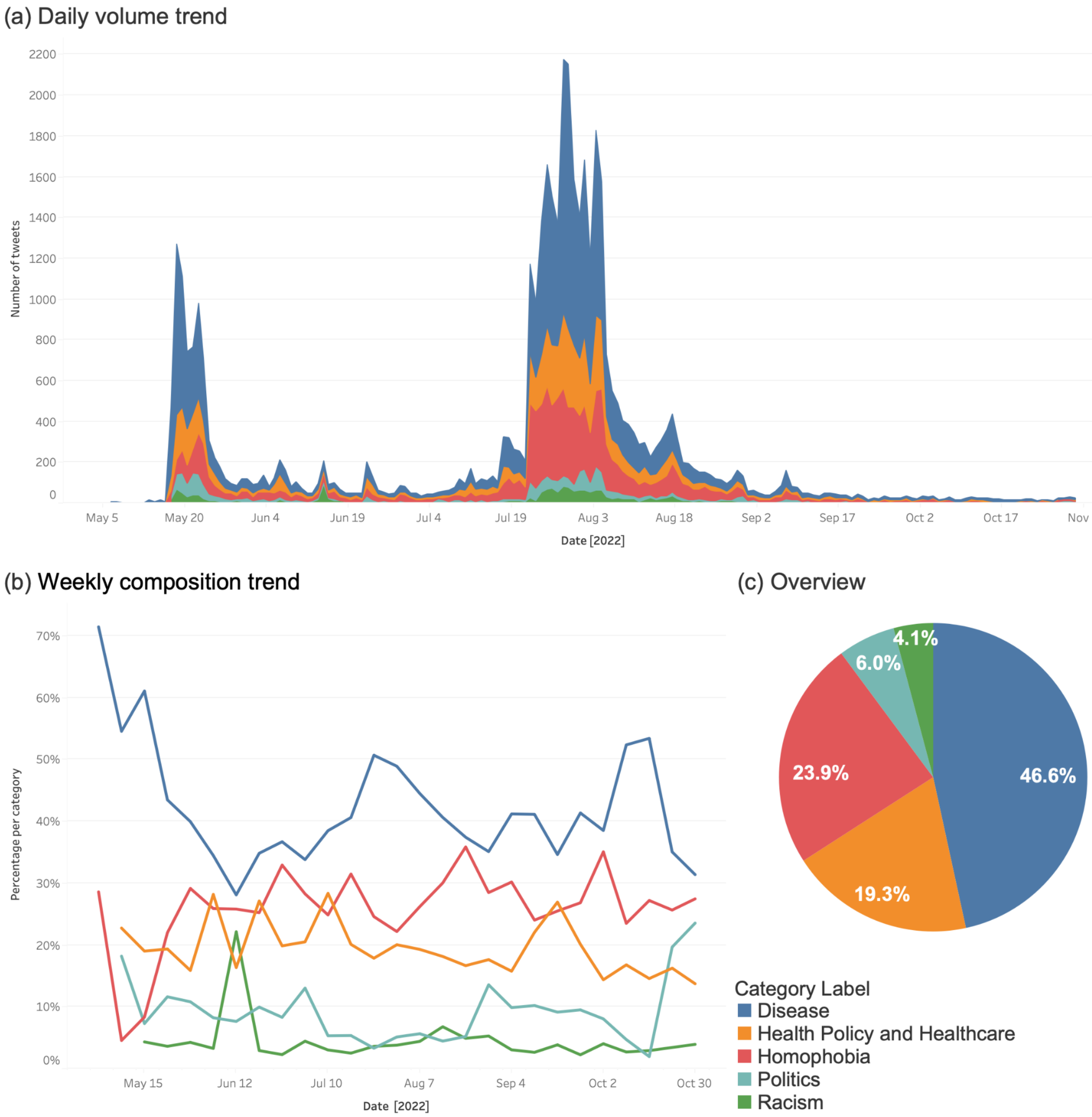

**Figure 4.** Toxicity categories overview and change over time

**Figure 4** shows the daily volume trend and weekly composition trend, as well as the composition overview of categories in toxic tweets during the Mpox outbreak (from May 6 to October 31, 2022). The overall trend indicates that a wide range of topics exist in the semantic dispersion of tweets: except for the first week when only two categories of discourse occur, we observe no discontinuity in any category, which shows strong topical diversity in toxic discourse.

Disease, Health Policy and Healthcare, and Homophobia are the dominant categories that have higher and comparatively stable compositions (often more than 10% of the overall discourse). At the same time, the Politics and Racism categories have lower daily volumes (often below 100 tweets) and weekly compositions (often below 10% of the discourse). These two categories also have more fluctuations in volume: there are continuous large discourse in the Politics category between May 20 and June 4 and in August as well as a composition increase towards the end of October; there is also large discourse in the Racism category around May 20, June 12, and between July 19 and August 23, with the cluster around June 12 comprising more than 20% of the discourse of that week.

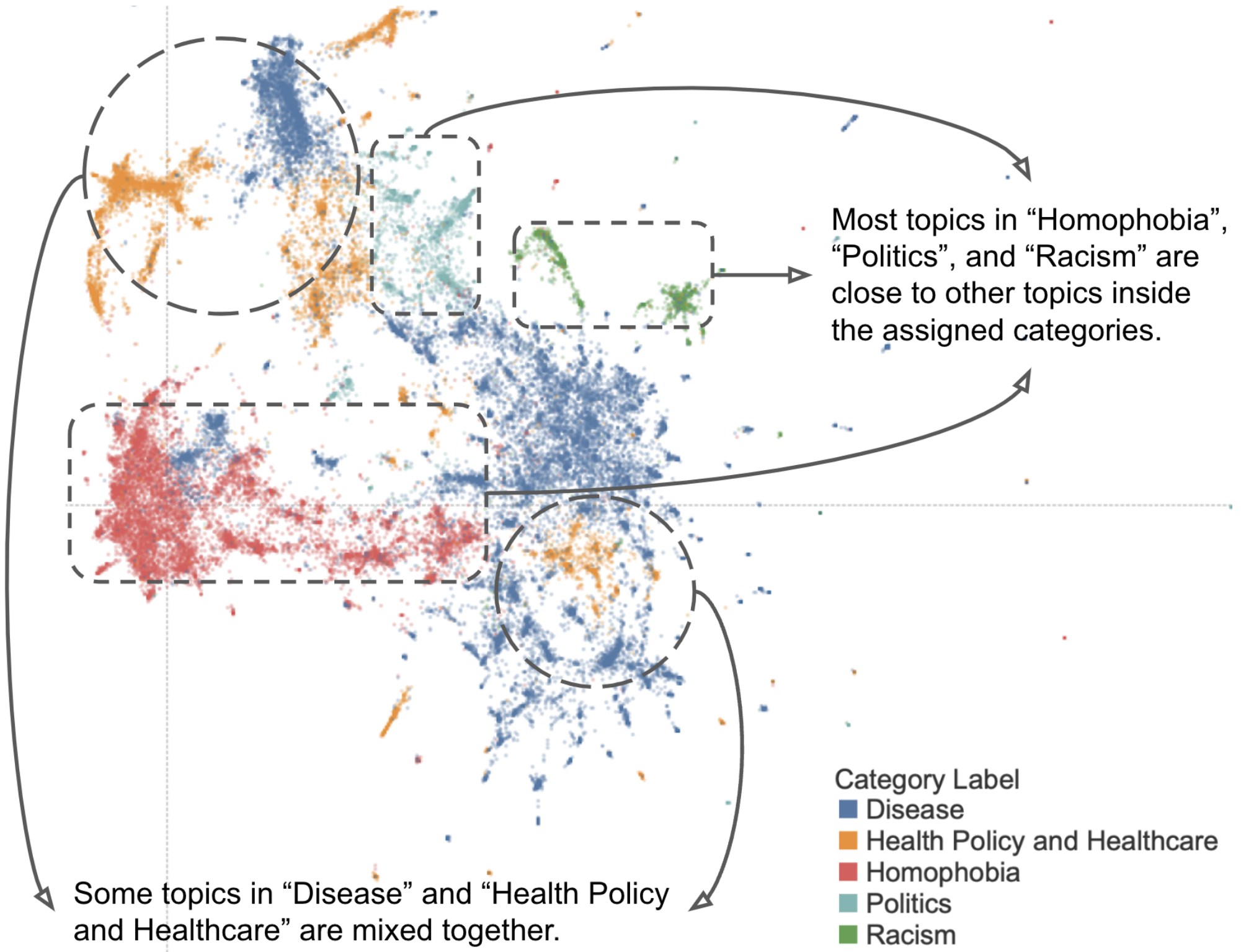

**Figure 5.** A two-dimensional visualization of toxic tweets with categories.

**Figure 5** demonstrates the inter-document and inter-category distances, based on the UMAP mapping of tweets to the semantic spaces of topics in a two-dimensional visualization [56], where the vertical and horizontal dashed lines are the axes of the two dimensions. The semantic space of toxic tweets are divided into several clusters, largely based on the summative categories of Twitter topics. The overall dispersion of the topics shows that the related topics belonging to the same category are close to each other, which indicates the comprehensiveness of the topic modeling and the categorization process. The collocation of topics further indicates internal relevance among toxic tweets of different topical focuses.

The Disease category and the Health Policy and Healthcare category are mixed together in multiple positions due to their topical relevance (highlighted in dashed circular frames in **Figure 5**). Practically, negative emotions on disease could either be amplified or mitigated depending on the effectiveness of health policies and services, which is the reason why the the topics in these two categories are visually close. It's often hard to talk about healthcare without talking about disease and illness, since they are inherently related. Homophobia, Politics, and Racism categories are respectively enclosed in connected areas, which demonstrate the different social political focuses in their individual discourse (highlighted in dashed rectangular frame in **Figure 5**). To better demonstrate the categorization criteria and contents in each category, we also provide example tweets in **Table 4** and more examples with notes in **Table 8** through **Table 12** in Appendix I.

**Table 4.** Example toxicity categories, topics, keywords, and tweets

| Category | Topic | Keywords | Example Tweet |
|---|---|---|---|
| **Disease** | 6 | scary_scared_shit_scaring | If monkeypox was a person lol I swear that face kills me but no more mate, he's scary as f*ck lol |
| **Health Policy and Healthcare** | 12 | health_emergency_outbreak_cdc | Hey CDC, F*ck You and your #monkeypox |
| **Homophobia** | 7 | sex_anal_transmitted_spread | Monkeypox is very serious, as serious as HIV for gay men having anal sex. The rest of us are Ok. Follow Health Guidelines… avoid anal sex with gay men. Listen to the science! Nuff said? #Canada |
| **Politics** | 11 | biden_ukraine_gates_f*ck | You bet your ass they will.. School shooting, monkeypox. Magically the story has changed away from Biden and his sh*tty gas prices, baby formula shortages, massive inflation, etc |
| **Racism** | 18 | n*gg*s_n*gg*_finna_yall | Laughing at someone catching monkeypox. You n*gg*s are lame frfr |

*Notes: 1) vowels in inappropriate words are masked; 2) emojis and some special characters are removed; 3) user names are removed; 4) some capital and lowercase letters; spaces, and punctuations are adjusted.*

# Information Diffusion Network

As mentioned in the Analytical Methods section, we focused on two relationships on Twitter: mentions and retweets. The mention network aims to reveal which accounts are frequently mentioned and so are encouraged to respond, while the retweet network aims to reveal which accounts diffuse toxicity. The research objective is to locate users who perpetuate or spread toxicity in the network. The mentions network (**Figure 6**) includes 17,437 vertices, 15,085 edges, and 3,628 connected components (i.e., clusters in the network). The average geodesic distance between two vertices is 8.041. The retweets network (**Figure 7**) includes 59,749 vertices, 62,493 edges, and 3,015 connected components. The average geodesic distance is 5.336. Overall, the mentions network has fewer vertices, edges, and connected components but a greater geodesic distance than the retweets network. This observation implies fewer interactions between users in the mentions network, possibly because those mentioned users did not respond to such toxic comments.

Regarding the users, we observed that a few users dominate the network, as centered in the cluster and surrounded by a large set of users in **Figure 6** and **Figure 7**. These "centered" users are frequently mentioned or retweeted by other users in the community and thus have the highest in-degree centrality. For some clusters, one user dominates the entire cluster (e.g., user 3 in cluster G3 in **Figure 6**). For other clusters, several users co-locate in the same cluster (e.g., user 2, user 9, and user 11 in cluster 2 in **Figure 6**), implying their tweets share similar outreaches and responses. We also observed that users in the mentions network are more dispersed than in the retweets network, as illustrated by a greater geodesic distance and a higher ratio of connected components divided by vertices. This reflects that a few users' tweets are repeatedly retweeted in the community, but users' mentions are arbitrary. However, many users were only mentioned or retweeted once according to both networks.

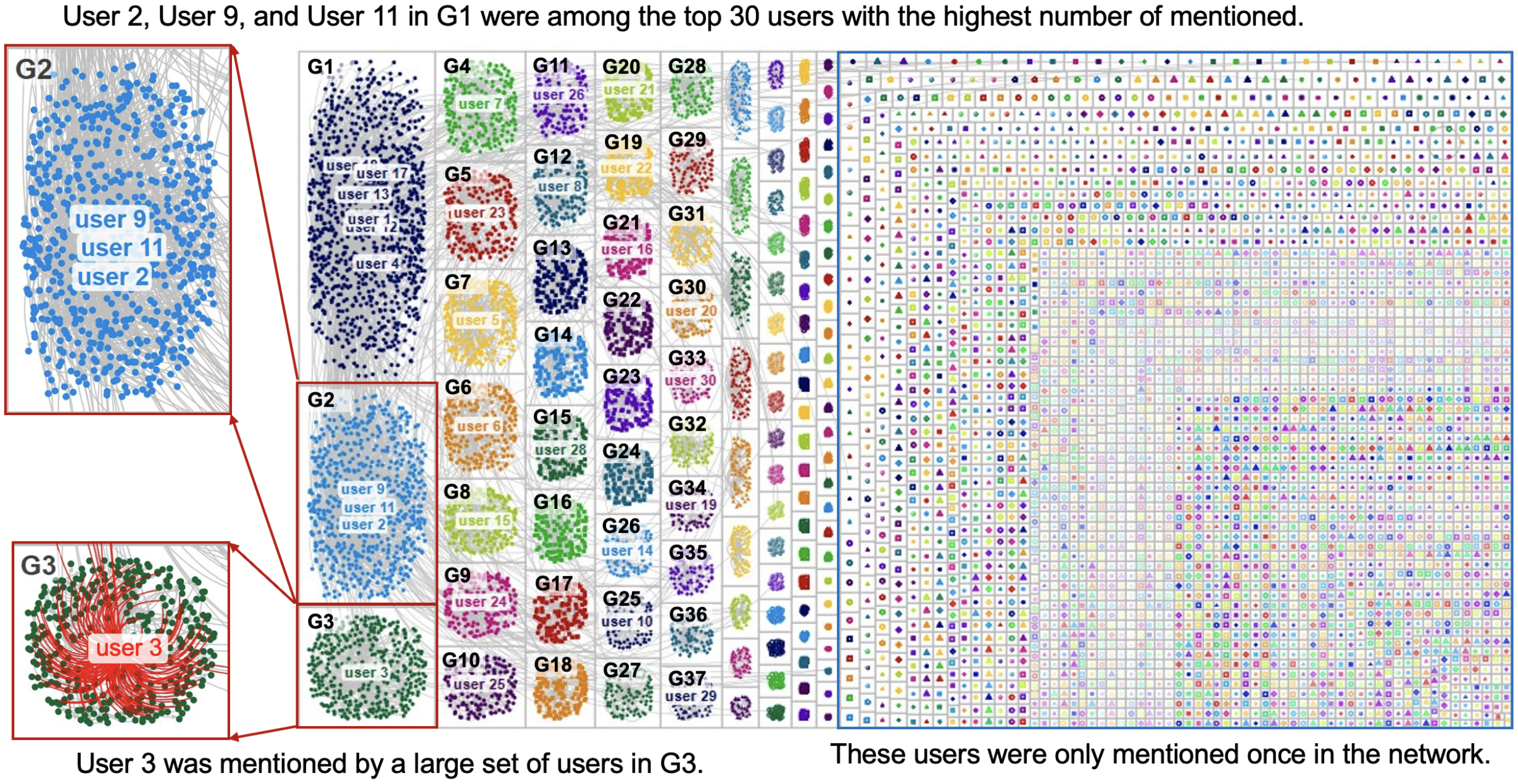

**Figure 6**. Social network of Twitter users based on mentions.

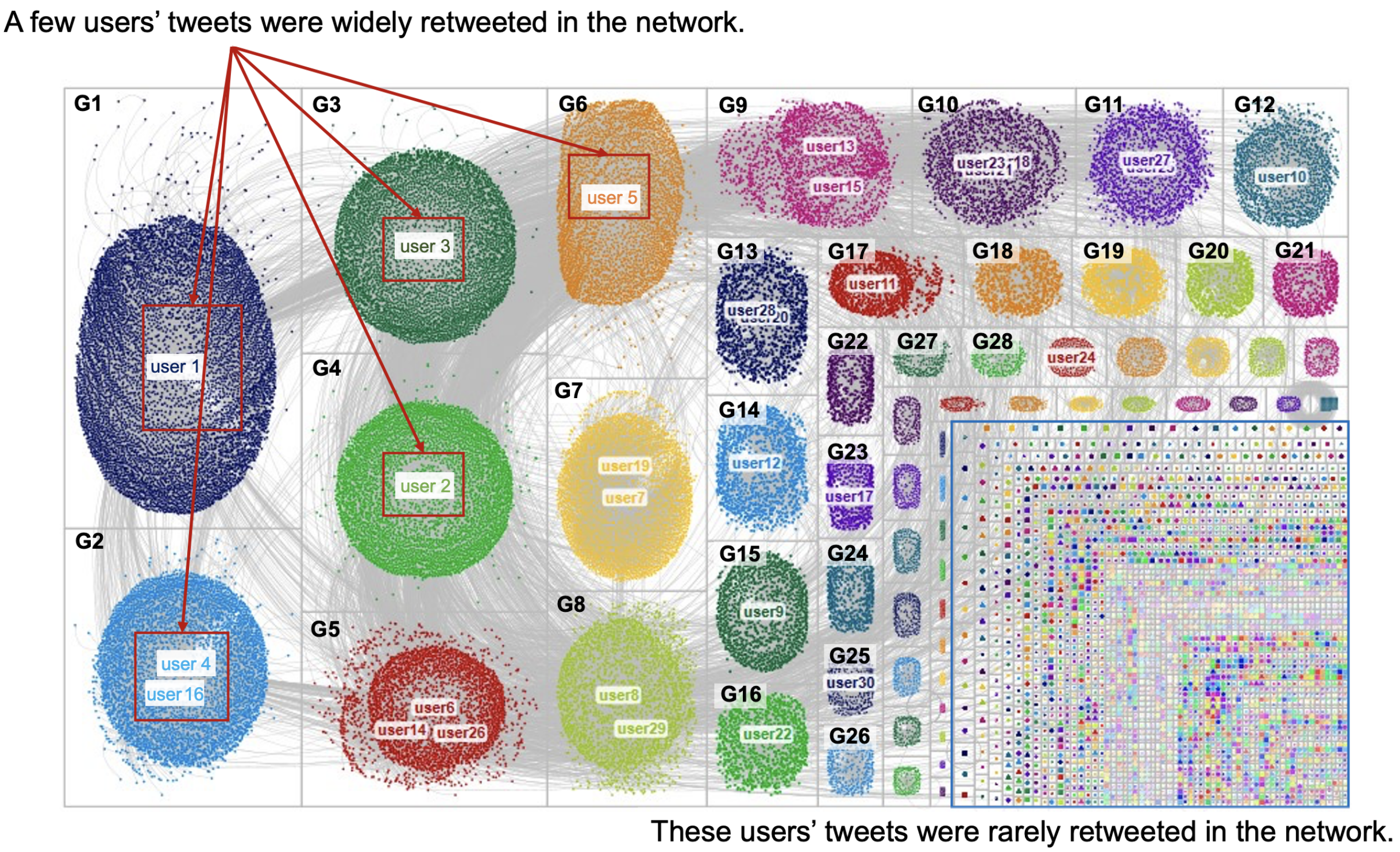

**Figure 7**. Social network of Twitter users based on retweets.

We further listed the top 30 users with the highest in-degree centrality with their account types in **Table 5** and **Table 6**. We used two attributes to describe a Twitter account. One attribute is the verification. A verified account may be an account of public interest, such as government agencies, politics, journalism, media, and influential public figures. For the verified accounts, we provide their usernames, while the usernames for non-verified accounts are masked. The other attribute is their account type. We manually interpreted each top user's account description and classified it into one of the following categories, as listed below:

- *Organization*: news media (org_media), government agencies (org_government)
- *Individual users*: politician (ind_politician), journalists (ind_journalist), high impact (ind_impact), other users (ind_other)

where we differentiated an individual account as an "high-impact user" or "other user" based on its number of followers; an account with more than 50,000 followers is identified as an "high-impact user."

**Table 5**. Top 30 users based on in-degree centrality in the *mentions* network.

| Username | In-degree | Cluster | Verified | Account type | Categories |
|---|---|---|---|---|---|
| User 1 (POTUS) | 229 | G1 | True | org_government | *D (68), H (47), O (67), P (46), R (2)* |
| User 2 (WHO) | 201 | G2 | True | org_government | *D (57), H (39), O (46), P (34), R (35)* |
| User 3 (RepMTG) | 155 | G3 | True | ind_politician | *D (49), H (9), O (67), P (31), R (1)* |
| User 4 (CDCgov) | 153 | G1 | True | org_government | *D (44), H (50), O (49), P (14), R (1)* |
| User 5 (TimRunsHisMouth) | 115 | G7 | True | ind_impact | *D (36), H (12), O (66), P (8), R (5)* |
| User 6 (FoxNews) | 111 | G6 | True | org_media | *D (37), H (14), O (33), P (18), R (13)* |
| User 7 | 80 | G4 | False | org_media | *D (26), H (12), O (31), P (10), R (1)* |
| User 8 (nypost) | 68 | G12 | True | org_media | *D (10), H (10), O (42), P (9), R (4)* |
| User 9 (CNN) | 67 | G2 | True | org_media | *D (19), H (5), O (34), P (9), R (5)* |
| User 10 | 63 | G25 | False | ind_impact | *D (17), H (4), O (37), P (6)* |
| User 11 (DrTedros) | 61 | G2 | True | ind_politician | *D (22), H (12), O (16), P (7), R (6)* |
| User 12 (JoeBiden) | 59 | G1 | True | ind_politician | *D (24), H (12), O (9), P (12), R (1)* |
| User 13 (CDCDirector) | 57 | G1 | True | org_government | *D (16), H (21), O (14), P (7), R (1)* |
| User 14 (SkyNews) | 50 | G26 | True | org_media | *D (21), H (2), O (13), P (5), R (12)* |
| User 15 (MrAndyNgo) | 46 | G8 | True | ind_journalist | *D (16), H (4), O (26), P (2), R (2)* |
| User 16 (JackPosobiec) | 44 | G21 | True | ind_politician | *D (13), H (4), O (24), P (3), R (1)* |
| User 17 (DrEricDing) | 41 | G1 | True | ind_impact | *D (20), H (8), O (15), P (3)* |
| User 18 (nytimes) | 40 | G1 | True | org_media | *D (11), H (6), O (18), P (7), R (1)* |
| User 19 (thehill) | 40 | G34 | True | org_media | *D (13), H (5), O (20), P (4), R (2)* |
| User 20 | 39 | G30 | False | org_media | *D (12), H (5), O (16), P (5), R (1)* |
| User 21 (Reuters) | 38 | G20 | True | org_media | *D (12), H (13), O (13), P (3), R (12)* |
| User 22 (Timcast) | 38 | G19 | True | ind_journalist | *D (18), H (4), O (13), P (1), R (2)* |
| User 23 (newsmax) | 37 | G5 | True | org_media | *D (18), H (4), O (11), P (2), R (1)* |
| User 24 (washingtonpost) | 36 | G9 | True | org_media | *D (7), H (3), O (24), P (1), R (1)* |
| User 25 (Scott_Wiener) | 35 | G10 | True | ind_politician | *D (4), H (5), O (22), P (3), R (1)* |
| User 26 (ZubyMusic) | 35 | G11 | True | ind_impact | *D (12), H (4), O (19), P (3)* |
| User 27 (BetoORourke) | 35 | G5 | True | ind_politician | *D (14), H (1), O (11), P (11)* |
| User 28 (bethanyshondark) | 34 | G15 | True | ind_impact | *D (15), H (5), O (15)* |
| User 29 (unusual_whales) | 33 | G37 | True | org_media | *D (10), H (7), O (16)* |
| User 30 (MattWalshBlog) | 33 | G33 | True | ind_impact | *D (5), H (5), O (25), R (1)* |

*Notes: 1) The number in parentheses in the Categories column indicate the occurrences of each category; 2) the sum of mentions in each category may be larger than the in-degree centrality in Table 2 since the in-degree centrality is computed based on the unique edges between two users.*

**Table 6**. Top 30 users based on in-degree centrality in the *retweets* network.

| Username | In-degree | Cluster | Verified | Account type | Categories |
|---|---|---|---|---|---|
| User 1 | 7,107 | G1 | False | ind_other | *D (1)* |
| User 2 (TimRunsHisMouth) | 5,038 | G4 | True | ind_impact | *D (1), H (1), O (1)* |
| User 3 | 4,325 | G3 | False | ind_impact | *D (1)* |
| User 4 | 4,072 | G2 | False | ind_impact | *D (1)* |
| User 5 | 3,222 | G6 | False | ind_impact | *D (1)* |
| User 6 | 2,258 | G5 | False | ind_impact | *H (1)* |
| User 7 | 1,947 | G7 | False | ind_impact | *D (1)* |
| User 8 | 1,000 | G8 | False | ind_impact | *H (1), R (1)* |
| User 9 | 897 | G15 | False | ind_impact | *D (1), O (1)* |
| User 10 (bahjarodriguez) | 804 | G12 | True | ind_impact | *D (1)* |
| User 11 | 599 | G17 | False | ind_impact | *D (1)* |
| User 12 (jennawadsworth) | 571 | G14 | True | ind_politician | *D (1)* |
| User 13 | 535 | G9 | False | ind_impact | *D (1)* |
| User 14 | 391 | G5 | False | ind_impact | *D (1)* |
| User 15 (DrEricDing) | 345 | G9 | True | ind_impact | *H (1)* |
| User 16 (AngryBlackLady) | 336 | G2 | True | ind_impact | *D (1)* |
| User 17 | 332 | G23 | False | ind_other | *H (1), O (1)* |
| User 18 | 291 | G10 | False | ind_impact | *H (1)* |
| User 19 | 285 | G7 | False | ind_impact | *D (1)* |
| User 20 | 284 | G13 | False | ind_other | *H (1)* |
| User 21 | 282 | G10 | False | ind_impact | *H (2)* |
| User 22 | 272 | G16 | False | ind_politician | *D (1), H (1), P (1)* |
| User 23 | 260 | G10 | False | ind_other | *H (1)* |
| User 24 | 259 | G29 | False | ind_impact | *D (1)* |
| User 25 (johncardillo) | 258 | G11 | True | ind_impact | *D (1)* |
| User 26 | 256 | G5 | False | ind_impact | *D (2)* |
| User 27 | 254 | G11 | False | ind_impact | *P (1)* |
| User 28 | 237 | G13 | False | ind_impact | *R (1)* |
| User 29 | 232 | G8 | False | ind_impact | *D (1)* |
| User 30 | 227 | G25 | False | ind_other | *O (1)* |

*Note: The number in parentheses in the Categories column indicate the occurrences of each category.*

We also noted a couple of observations regarding the top users in the networks. For the mentions network, 90% (27 out of 30) of the top-mentioned users have verified accounts, primarily news agencies, government portals, politicians, and independent high-impact users. In particular, the most frequently mentioned accounts are from government portals or politicians. By contrast, for the retweets network, most top users are non-verified accounts (24 out of 30), and they are independent influencers (23 out of 30) with large followings (i.e., more than 50,000 followers). There is also a clear distinction regarding the organization accounts between mentions and retweets networks. More than half of the most frequently mentioned accounts are from organization accounts, but none are among the top users in the retweets network.

Regarding the categories, these top users were frequently mentioned in tweets relative to Disease, Health Policy and Healthcare, and Homophobia, but comparatively less mentioned in the categories of Politics and Racism. Based on the retweets network, those tweets discussing Disease and Health Policy and Healthcare were most likely to obtain attention from the online community, but the other three categories did not. We observed that the categories of Homophobia, Politics, and Racism were more likely to be shorter- and more locally-lived than those Disease and Health Policy and Healthcare toxicity since those topics rarely appeared among the top retweeted users. Toxic tweets in the categories of Homophobia and Racism did not receive much attention, given that categories of “O” and “R” were rarely mentioned in **Table 6**.

# Discussion

Online toxicity is widespread during health crises, with many individuals spreading misinformation, fear, and hatred [8,9]. This can undermine public health communication efforts and lead to confusion and anxiety among the public. The discussion of toxic narratives during the 2022 Mpox outbreak is an example of controversy of public communication during health crises. Building on prior works that leverage either topic modeling or network analysis techniques [67–69], our study further demonstrates the value of combining topical and network analyses to understand emerging social issues and crises. By examining the topical dynamics, we were able to uncover the prevalent themes in the toxic discourse during the 2022 Mpox outbreak and observe their temporal shifts. Network dynamics revealed the key users and their roles in propagating toxicity, suggesting that addressing these high-impact users and their narratives could be crucial for effective crisis communication and policy decision-making. Our findings highlight the importance of monitoring and addressing online toxicity to foster a more inclusive and constructive public dialogue during health emergencies.

We also adapted the Rabat Plan of Action analytical framework for hate speech analysis to study online toxicity during the Mpox outbreak. This framework takes into account context, extent, content, speaker, and intent. By examining the context of the discussions, including the events that led to the discourse, the extent of toxic comments and hate speech, the content and themes, the speakers involved, and the intent behind the messages, the adapted analytical framework provided a comprehensive understanding of the toxicity landscape in the Mpox scenario.

## Toxicity aboutness reveals an infodemic: negative feelings, political unrest, and weaponized stigma

Topical dynamics summarize temporal content popularity and provide extended context of social issues in the Mpox health crisis. In this sense, the understanding of context and content of toxicity are both representing the extremes of public opinion, respectively from event or topical trend perspectives, that are mutually beneficial in profiling the problems in health communications during the 2022 Mpox outbreak. By examining groups of related topics, we can grasp an overarching view of the main subjects being discussed. In other words, we can identify the primary categories that highlight the most commonly mentioned topics. By analyzing temporal topical swifts, we further understand when the topical discourses, especially their peaks, occur in each of the categories. Such summary of contents can reveal what categories of topics are discussed together, which quantitatively demonstrate public opinions around context and facilitate a multifaceted understanding of toxic contents in this health crisis. In particular, there are three outstanding problems associated with online toxicity: negative feelings, political unrest, and weaponized stigma.

Negative feelings during the 2022 Mpox outbreak often came out of emotions towards the disease or how the disease was dealt with. There are negative feelings such as fear and anger due to the actual or imagined physical symptoms caused by Mpox or negativity because of mental anxiety. Specific to the 2022 Mpox outbreak, there are also negative feelings carried on from COVID-19, since some initial public health guidance were similar (e.g., vaccination, wearing masks, and self-quarantine). Thus, similar to COVID-19, people can also be unhappy about how the health emergency is dealt with by the health authorities and healthcare providers. These negative feelings are not produced in isolation: Twitter users read and watch news from different media platforms and share their ideas on the platform. When some personal beliefs, which might not be scientifically mature, are put together with the practical inconvenience, the public uses toxicity to express their negativity towards the disease and the health services they received.

Political unrest during the 2022 Mpox outbreak was fueled by divisive reactions from politicians and the public, leading to the spread of conspiracy theories and misinformation. Disagreements over health policies, allocation of resources, and the overall handling of the outbreak often manifested in toxic discourse, further polarizing society. For example, some Twitter users propagated unfounded claims that the Mpox outbreak was a result of a laboratory leak or a government conspiracy, which led to increased distrust in the authorities and healthcare providers. This toxic environment can be understood through the lens of psychological factors such as fear, uncertainty, and a tendency towards confirmation bias, where individuals are more likely to believe and spread information that aligns with their pre-existing beliefs and fears. Societal factors, such as political polarization and a general erosion of trust in public institutions, also played a critical role in amplifying toxic behavior online, as people sought out and shared content that validated their anxieties and skepticism.

Weaponized stigma became another significant issue during the outbreak, as incidents of attacks towards minority groups on the basis of (perceived) sexuality, gender identity, and race, were reported. This stigmatization was often rooted in misinformation and fear, with people associating certain groups with the spread of the disease or accusing them of not following public health guidelines. Online toxicity facilitated the perpetuation of these stigmatizing narratives, further marginalizing these communities and exacerbating existing social divisions. Psychological factors such as xenophobia, scapegoating, and the need to find a tangible source of blame during a crisis contributed to the spread of these harmful narratives. Societal factors, including systemic discrimination and historical prejudices, were also at play, as these pre-existing biases were amplified in the digital space, leading to more virulent expressions of hate and intolerance. Understanding these dynamics is crucial for developing strategies to mitigate online toxicity and support affected communities.

## Toxicity diffusion suggests improvements in health communication: influential users should respond to and counter toxicity

Network dynamics reveal frequent speakers and intents and suggest priorities in public health communications and health policy. Based on the analytical framework, we used social network theory to calculate the in-degree and betweenness centrality. Through this approach, we were able to identify influential speakers and mentioned users, as well as gain insights into what they said and the likelihood of responses in online communities. Our analysis shows several observations regarding speakers and intents that are worth discussing.

In our analysis of speakers, we discovered that a few non-verified yet influential users dominated the retweets network. Their tweets garnered broad attention from the online community and resonated with many others who shared the same opinion regarding disease- and health-related negativity. By contrast, users who were mentioned in these toxic messages appeared to be scattered and chosen randomly, and they seldom responded to the negativity. Our findings suggest that toxic information, regardless of its intent, typically does not elicit responses from those who are mentioned.

For those most frequently mentioned users, our analysis revealed that verified government channels, news agencies, and politicians dominated the top-mentioned list. This finding highlights the importance of government and health agents being aware of toxic information and taking appropriate action. For example, some users expressed concerns on Twitter about the transmission routes of a disease, albeit in a toxic manner. To help limit such toxicity, we suggest that public health organizations such as the WHO and CDC should inform the public about the transmission routes of the disease and the severity of the health crisis. This underscores the need for timely and effective communication from official sources in response to public concerns.

Upon analyzing intents in the retweets network, we observed that attributions of diseases to homophobia and racism were not frequently mentioned by the top speakers. While one interpretation could be that most online users view such attributions as malicious during public health crises, another perspective is that these top speakers, who are already prominent and attract attention regardless of their content, might not engage in such rhetoric. If these influential speakers had used this kind of rhetoric, it's possible it would have still received significant engagement. In contrast, tweets related to disease- and healthcare policy-related negativity were more generalized among a broader set of users and were retweeted for a more extended period. Again, our findings highlight the importance of focusing on accurate and relevant information during public health crises, as misinformation or toxic narratives can be quickly dismissed by online users. We suggest public health agencies prioritizing accurate information dissemination, which can help combat the spread of harmful narratives and promote healthy dialogue and public understanding of health-related issues.

After analyzing intents in the mentions network, we found that verified users were primarily mentioned in topics related to disease-related negativity, health policy negativity, and homophobia. Our findings reveal widespread dissatisfaction regarding health policies for Mpox and concerns about the severity of the disease outbreak. This highlights the urgent need for government or health channels to release transparent and reliable information to address public concerns. Our findings also imply that some users might misunderstand the disease transmission or use Mpox to stigmatize homosexuality. This indicates a need for relevant agencies to take immediate action to interrupt the dissemination of toxic information. By doing so, we can mitigate the influence of toxicity and reduce the harming of historically marginalized groups such as the LGBTQ+ community.

## Comparison to Prior Work

Prior research has demonstrated that NLP techniques, such as topic modeling [67] and text classification [70], could be useful for uncovering online toxicity and hate speech during health crises. In the broader context of health communication, our study highlights several significant implications. One key implication involves the use of NLP and computational techniques to analyze social media narratives. Our research further shows that combining topic modeling and network analysis can provide a nuanced understanding of the online toxic narratives and their dissemination.

Next, our topic modeling results suggest common underlying causes of toxicity and hate speech on social media, including emotional, political, and stigmatizing responses. Toxic narratives by negative feelings have been widely reported in previous studies [71,72], often linked to policies like lockdowns and mask mandates during COVID-19 [73]. Additionally, political unrest and misinformation exacerbate these emotions, with conspiracy theories and distrust in authorities intensifying toxic narratives. This pattern, observed during the Mpox outbreak, was also prevalent in past pandemics such as COVID-19 [68]. Moreover, the stigmatization of certain groups, such as the gay community during the Mpox outbreak, mirrors the scapegoating and xenophobia seen in previous crises. For instance, prior studies reported widespread anti-Asian sentiment on social media during COVID-19 [74,75]. Understanding these causes in terms of misinformation, distrust of government or health agencies, and societal biases on certain groups, is crucial for developing strategies to mitigate the harmful impact of online toxicity during health crises.

Our network analysis highlights the crucial need to engage with influential users and address key narratives to mitigate the spread of online toxicity during health crises. Consistent with previous research [33], our findings suggest that official sources—such as government agencies and healthcare organizations—must prioritize timely, transparent, and accessible communication across major social media platforms, as these entities are frequent targets of toxic discourse. While earlier studies have found that right-wing sources are often associated with higher levels of toxicity and scientific sources with lower levels [68], our research indicates that general users, rather than verified accounts, frequently lead and propagate toxic narratives. This underscores the importance of not only monitoring influential users but also addressing the broader network of general users who contribute to the dissemination of harmful content.

## Limitation and outlook

One limitation could result from the use of Perspective API. For example, Perspective API might incorrectly identify the toxicity if a tweet's toxic words or patterns do not appear similar to its training samples. In particular, social media language is informal, and Perspective API might not be able to identify internet slang or abbreviations correctly. In addition, the threshold selection may impact the toxicity results. A smaller threshold increases the likelihood of identifying a toxic tweet. However, it simultaneously increases the number of false positives (i.e., a non-toxic tweet is identified as toxic) [10].

Another limitation relates to the types of toxicity used. Online toxicity may include both emphasis of emotions (e.g. fear and anxiety) towards the crisis and attacks of associated minority groups (e.g,. Asian Americans for the COVID-19 pandemic and LGBTQ+ people for the 2020 Mpox outbreak). There is no strict boundary between the users who spread these two types of toxicity, nor between the language that they use. For example, although an essential proportion of the toxicity here is indeed attacking LGBTQ+ or African and African American communities, it is hard to easily tell if some other tweets are using toxic words to highlight the inequity or discrimination and condemn the identity attacks. Future work could include an analysis of non-toxic discourse to serve as a comparison to facilitate the overall understanding of toxic tweets about Mpox. By examining the overlap between the discourse in toxic versus non-toxic discourse, especially the influential users behind it, we can better understand the extent to which these users are involved in combating misinformation and toxic speech. This comparative analysis would provide a more comprehensive depiction of user discourse on social media, with a focus on showcasing the difference between emphasis and toxicity.

In addition, algorithmic bias is an important consideration in our study, particularly concerning the algorithms employed for data analysis, such as BERT and UMAP. BERT, a transformer-based model, is pre-trained on large text corpora which may contain inherent biases reflecting societal stereotypes and prejudices. These biases can influence the model's understanding and representation of language, potentially skewing the identification and clustering of topics. Similarly, UMAP, a dimensionality reduction technique, might introduce biases through its assumptions about data structure and the preservation of local and global relationships within the dataset. These algorithmic biases can impact the study's findings by potentially misrepresenting the semantic relationships and topic distributions within the toxic discourse, leading to conclusions that may not fully or accurately reflect the underlying data. Acknowledging these biases is crucial, and future work should focus on employing bias mitigation strategies, such as algorithmic auditing and using debiased training datasets, to enhance the fairness and accuracy of the analysis.

At the same time, toxicity can look differently in different parts of the world, among different cultural groups, and in different languages. While our analysis focuses on the English-speaking countries, we need to be aware that many influenced populations are not covered by the tweets we collected. Our approach may inherently overlook the perspectives and nuances present in non-English tweets, which may result in a cultural bias. We chose English content primarily due to the availability of robust language processing tools and resources for English, which facilitates more reliable analysis. We recognize that this focus might limit the generalizability of our findings to non-English speaking audiences and have listed this as limitation in discussion. The discourse around public health crises, like the Mpox outbreak, is multifaceted and culturally dependent. Non-English content may reveal different concerns, misinformation patterns, and public reactions that our study does not capture. As such, we could investigate toxicity with more granularity in the future and characterize it with regard to attack versus emphasis, as well as demography factors including language, country, and culture.

## Conclusions

Toxic online discourse can have detrimental impacts on public health crises. In this study, we collected tweet data during the 2022 Mpox outbreak and analyzed toxicity from multiple dimensions, including context, extent, content, speaker, and intent. To better understand toxic dynamics on Twitter, we utilized BERT-based topic modeling and social network community clustering techniques.

The temporal discourse analysis revealed that toxic tweets during the outbreak covered a diverse range of topical categories. The predominant topics were toxicity on disease, health policy and healthcare, and homophobia. While toxicity related to politics and racism had lower daily volumes, they reached respective peaks when triggering events happened. On the other hand, verified government channels, news agencies, and politicians were among the top-mentioned users in the social network and were primarily associated with the categories of Disease, Health Policy and Healthcare, and Homophobia. Meanwhile, a few non-verified but influential users posted tweets that received high volumes of retweets, and tweets related to homophobia, politics, and racism were more likely to be shorter and have a local impact.

As such, to mitigate online toxicity of Mpox or similar infodemics, public health authorities should leverage advanced natural language processing tools, such as sentiment analysis and toxicity detection algorithms, to identify and address harmful content in real time. Additionally, digital literacy campaigns can educate the public about the dangers of misinformation and the importance of respectful online discourse. Establishing rapid response teams comprising public health experts, communication specialists, and community leaders can help counteract false narratives and provide accurate information swiftly. Finally, fostering partnerships with influential social media figures and organizations can amplify positive messages and mitigate the spread of toxic content.

To summarize, the topical dynamics revealed that Twitter users were expressing negativity and making controversial remarks about the Mpox outbreak, indicating a worsening of political unrest and the increased weaponization of stigma during the corresponding infodemic. The network dynamics highlight the need for government and health agencies to release transparent and reliable information to address public concerns. Overall, our study demonstrates a workflow that combines topical and network analyses to understand emerging social issues and crises. Our findings emphasize the importance of proactive measures needed from government and health agencies to combat harmful narratives and promote accurate information during public health crises.

# Funding Statement

This material is supported by the Carnegie Endowment for Faculty Development at UMSI.

# Data Availability

The Twitter data utilized in this study is subject to Twitter’s Developer Agreement and Policy. As such, raw data consisting of tweets and user information cannot be shared directly due to privacy and data protection regulations. The processed data files are available upon reasonable request.

# Author Contributions

The authors' contributions are as follows. Lizhou Fan: Conceptualization, Methodology, Data Curation, Formal Analysis, Writing - Original Draft, Writing - Review & Editing. Lingyao Li: Conceptualization, Methodology, Formal Analysis, Writing - Original Draft, Writing - Review & Editing. Libby Hemphill: Conceptualization, Writing - Review & Editing, Project Administration, Resources, Supervision.

# Declaration of Generative AI Use

The authors declare no direct use of generative AI in authoring the manuscript. Third-party grammar check applications we used (application name: Grammarly).

# Conflicts of Interest

None declared.

# Abbreviations

BERT: bidirectional encoder representations from transformers
BERTopic: a topic modeling technique that leverages BERT
COVID-19: coronavirus disease, an infectious disease caused by the (Severe Acute Respiratory Syndrome Coronavirus-2) SARS-CoV-2 virus
Mpox: formerly known as monkeypox, is an infectious disease caused by the monkeypox virus.
NLP: natural language processing
SBERT: or Sentence-BERT, sentence embeddings using Siamese BERT networks

# Appendices

## I. Notes on Topic Modeling

We provide more examples for contextual understanding of the five categories identified through topic modeling and discourse analysis. vowels in inappropriate words are masked.

**Table 7.** Example topics and tweets of **Disease**

| Topic | Keywords | Note | Example |
|---|---|---|---|
| 15 | ugly_looks_look_sh*t | Negativity because of physical symptom (disgust) | That MonkeyPox sh*t look so nasty |
| 6 | scary_scared_shit_scaring | Negativity because of mental anxiety (scarcity) | If monkeypox was a person lol I swear that face kills me but no more mate, he's scary as f*ck lol |
| 1 | covid_sh*t_yall_monkeypox | Negativity carried on from related diseases | First Covid now monkeypox? This sh*ts bananas |
| 2 | bullshit_monkeypox_f*ck_monkey | Denial of disease or emergency | i refuse to learn anything about 'monkeypox' f*ck you |

**Table 8.** Example topics and tweets of **Health Policy and Healthcare**

| Topic | Keywords | Note | Example |
|---|---|---|---|
| 12 | health_emergency_outbreak_cdc | Declaration and development | Hey CDC, F*ck You and your #monkeypox |
| 10 | vaccine_vax_monkeypox_im | Precaution (vaccine) | Monkeypox vaccine is not the same, you mass media drinking wh*r*. |
| 27 | mask_masks_wear_wearing | Precaution (mask) | People masking up for monkeypox is the ultimate stupidity. |
| 34 | lockdown_lock_sh*t_lockdow | Precaution (lockdown) | As proved lockdowns we're f*ck*n pointless. There was no need for them but cos folk are so gullible; stupid they actually complied without asking questions. Now u know not to comply when there's future lockdowns regarding this Monkeypox which there will. Wake the f*ck up |
| 32 | school_year_kids_senior | Daily life impact (schooling) | people concerned kids are getting monkeypox at school ≠ satanic panic. f*cking idiots co-opting terms to make legitimate concerns look crazy. |
| 36 | gym_crib_going_im | Daily life impact (exercise) | No gym, clubs, or link-ups for awhile for me. This monkeypox sh*t has me shook all over again. |

**Table 9.** Example topics and tweets of **Homophobia**

| Topic | Keywords | Note | Example |
|---|---|---|---|
| 7 | sex_anal_transmitted_spread | Medical scapegoating | Monkeypox is very serious, as serious as HIV for gay men having anal sex. The rest of us are Ok. Follow Health Guidelines… avoid anal sex with gay men. Listen to the science! Nuff said? #Canada |
| 16 | ass_eating_butt_asshole | Sexually explicit remarks | Outside eating random ass is insane but eating random ass when monkeypox is a thing is NUTS |

**Table 10.** Example topics and tweets of **Politics**

| Topic | Keywords | Note | Example |
|---|---|---|---|
| 11 | biden_ukraine_gates_f*ck | Conspiracy theory | You bet your ass they will.. School shooting, monkeypox. Magically the story has changed away from Biden and his sh*tty gas prices, baby formula shortages, massive inflation, etc |
| 35 | greene_taylor_marjorie_shes | Political remarks | Marjorie Taylor Greene has just humiliated herself once again as she is thick and stupid she doesn't know what the hell she is talking about ok don't listen to her monkeypox is not a sexual transmitted disease there is no evidence of this and kids have not caught it |

**Table 11.** Example topics and tweets of **Racism**

| Topic | Keywords | Note | Example |
|---|---|---|---|
| 18 | n*gg*s_nigg*_finna_yall | Racial slur towards African American | Laughing at someone catching monkeypox. You n*gg*s are lame frfr |
| 29 | racist_black_africa_white | Remarks of racism | B*tch I thought monkeypox was some racist shit klan whites were trying to popularize |

## II. Method Notes on Network Analysis

We leveraged social network theory to discover toxic communication patterns during the Mpox crisis. **Figure 8** illustrates the relationships among Twitter users and the degree centrality theory: a. Two ways of communication on Twitter. b. An illustration of a Twitter network.

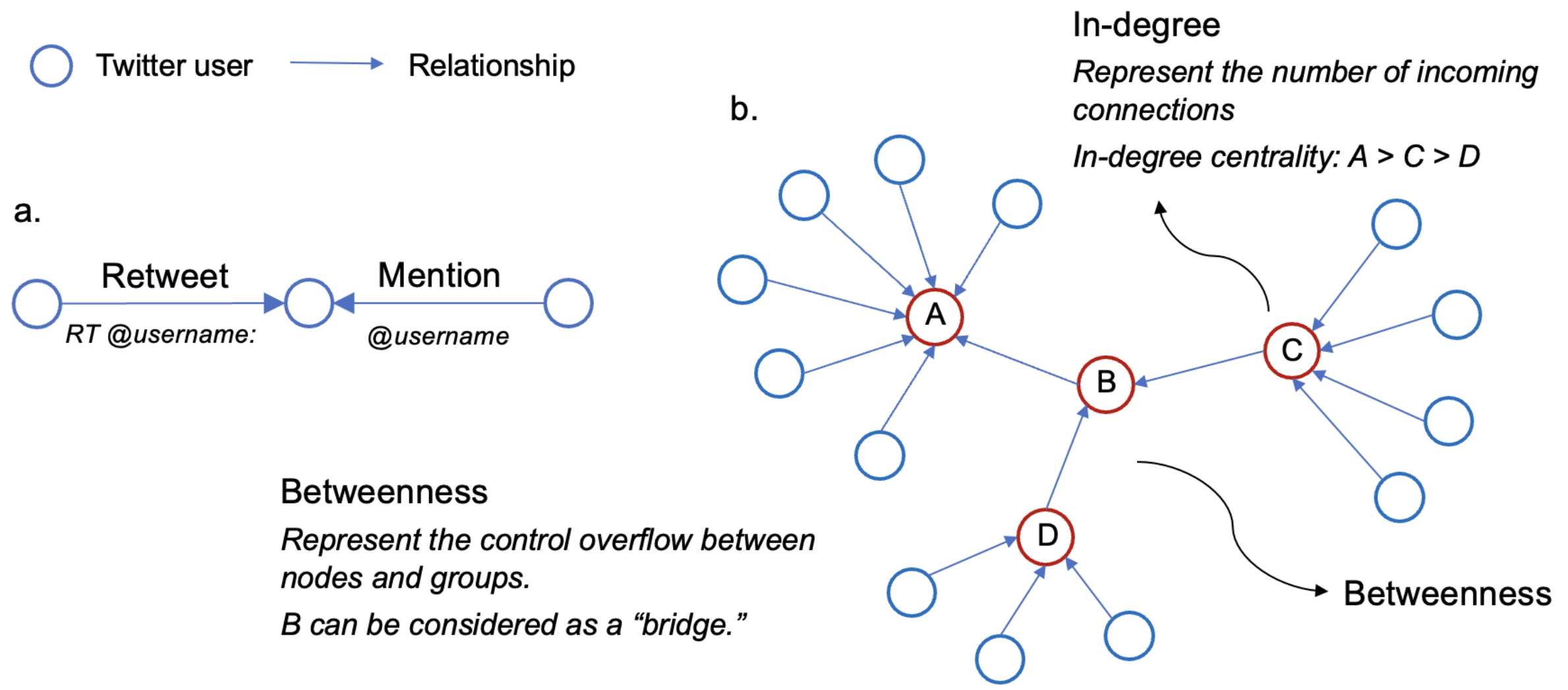

**Figure 8.** Relationships among Twitter users and the degree centrality theory

In addition to the top users based on degree-centrality, we have also investigated the user ranking via betweenness centrality, as attached in the appendix. The top users remain largely the same as using the in-degree centrality: 26 of the top 30 users remain the same in the retweets network. A user with a higher betweenness centrality in the network implies that more information passes through this user. In this context, those top users who communicated toxicity also played a critical role in communicating other toxic information to others. Thus, it is sufficient to only use degree-centrality top users for our analysis.

**Table 12.** Top 30 users based on betweenness centrality in the retweets network.

| Username | Betweenness | Cluster | Verified | Account type |
|---|---|---|---|---|
| User 1* | 963,382,779.9 | G1 | False | ind_other |
| User 2* | 876,809,948.9 | G2 | False | ind_impact |
| User 3* | 764,206,632 | G4 | True | ind_impact |
| User 4* | 509,923,912.6 | G3 | False | ind_impact |
| User 5* | 352,379,040.6 | G6 | False | ind_impact |
| User 6 | 237,515,582.5 | G2 | False | ind_other |
| User 7* | 232,910,047.8 | G5 | False | ind_impact |
| User 8* | 205,528,733.5 | G7 | False | ind_impact |
| User 9 | 140,960,282.8 | G3 | False | ind_other |
| User 10 | 113,780,200 | G6 | False | ind_other |
| User 11* | 113,734,640.7 | G15 | False | ind_impact |
| User 12* | 98,522,889.35 | G12 | True | ind_impact |
| User 13 | 92,042,632.22 | G4 | False | ind_other |
| User 14* | 90,506,669.18 | G8 | False | ind_impact |
| User 15* | 84,159,899.17 | G9 | False | ind_impact |
| User 16* | 77,754,330.19 | G14 | True | ind_politician |
| User 17 | 73,277,436.47 | G2 | False | ind_other |
| User 18* | 68,551,204.88 | G2 | True | ind_impact |
| User 19* | 67,826,922.9 | G17 | False | ind_impact |
| User 20 | 57,119,996.73 | G7 | False | ind_other |
| User 21* | 51,240,533.17 | G5 | False | ind_impact |
| User 22* | 45,905,200.3 | G23 | False | ind_other |
| User 23* | 45,481,426.53 | G19 | False | ind_impact |
| User 24 | 43,931,889.3 | G6 | False | ind_other |
| User 25 | 43,887,495.78 | G41 | False | ind_other |
| User 26* | 40,839,333.65 | G16 | False | ind_politician |
| User 27* | 39,969,804.56 | G9 | True | ind_impact |
| User 28* | 38,173,697.98 | G10 | False | ind_impact |
| User 29 | 36,740,190.94 | G10 | False | ind_other |
| User 30* | 33,068,415.68 | G20 | False | ind_impact |

*Note: in the table, the symbol "*" indicates that the user also ranked as one of the top 30 users based on the in-degree centrality*.